\documentclass{aa}  

\usepackage{graphicx}

\usepackage{txfonts}
\begin{document} 

\titlerunning{The quest for planets around subdwarfs and white dwarfs}
   \title{The quest for planets around subdwarfs and white dwarfs from \textit{Kepler}
space telescope fields: Part I}
   \subtitle{Techniques and tests of the methods}

 \author{ J.~Krzesinski\inst{1}\and A.~Blokesz\inst{2}\and M.~Siwak\inst{3}\and G.~Stachowski\inst{2}}

   \institute{Astronomical Observatory, Jagiellonian University, 
        ul. Orla 171, PL-30-244 Krakow, Poland
        \\
               \email{jk@oa.uj.edu.pl}, \email{abl@astro.as.up.krakow.pl}
             \and
             Mt. Suhora Observatory, Pedagogical University of Cracow,
             ul. Podchor\c{a}\.{z}ych 2, 30-084 Cracow, Poland
             \and
             Konkoly Observatory, Research Centre for Astronomy and Earth Sciences,
             Konkoly-Thege Mikl\'{o}s ut 15-17, 1121 Budapest, Hungary
             }
   \date{Received 2020}

 
  \abstract
   {
In this study, we independently test the presence of an exoplanet around the 
binary KIC\,9472174, which is composed of a red dwarf and a pulsating type B 
subdwarf. We also present the results of our search for Jupiter-mass objects 
orbiting near to the eclipsing binary KIC\,7975824, which is composed of 
a white dwarf and type B subdwarf, and the pulsating white dwarf KIC\,8626021.
   }
   {
The goal is to test analytical techniques and prepare the ground for a larger 
search for possible substellar survivors on tight orbits around post-common 
envelope binaries and stars at the end of their evolution, that is, 
extended horizontal branch stars and white dwarfs.
We, therefore, mainly focus on substellar bodies orbiting these stars within the range of 
the host's former red-giant or asymptotic-giant phase envelopes. Due to the methods we use,
the quest is restricted to single-pulsating type B subdwarf and white dwarf stars and 
short-period eclipsing binaries containing a white dwarf or a subdwarf component.      
   }
   {                    
Our methods rely on the detection of exoplanetary signals hidden in photometric time series 
data from the $\it{Kepler}$ space telescope, and they are based on natural clocks within the 
data itself, such as stellar pulsations and eclipse times. The light curves are analyzed 
using Fourier transforms, time-delays, and eclipse timing variations.  
   }
   {
Based on the three objects studied in this paper, we demonstrate that 
these methods can be used to detect giant exoplanets orbiting around pulsating white 
dwarf or type B subdwarf stars as well as short-period binary 
systems, at distances which fall within the range of the former red-giant envelope 
of a single star or the common envelope of a binary.
Using our analysis techniques, we reject the existence of a Jupiter-mass
exoplanet around the binary KIC\,9472174 at the distance and orbital 
period previously suggested in the literature. We also found that the eclipse timing 
variations observed in the binary might depend on the reduction and 
processing of the  $\it{Kepler}$ data. The other two objects analyzed in this work do not 
have Jupiter mass exoplanets orbiting within 0.7\,--\,1.4\,AU from them, or larger-mass 
objects on closer orbits (the given mass limits are minimum masses).    
   }
   {
Depending on the detection threshold of the time-delay method and the inclination of the exoplanet orbit toward the observer, data from the primary $\it{Kepler}$ mission allows for the detection of bodies with a minimum of $\sim1$\,Jupiter-mass orbiting these stars at $\sim$1\,AU, while data from the K2 mission extends the detection of objects with a minimum mass of $\sim7$ Jupiter-mass on $\sim0.1$\,AU orbits. The exoplanet mass and orbital distance limits depend on the length of the available photometric time series.
   }


   \maketitle

%

\section{Introduction}

Recent studies suggest that, statistically, most of the stars in our Galaxy are orbited 
by exoplanets \citep{Cassan2012}. While the vast majority of them have been detected
around main sequence stars, the number of detected exoplanets around subdwarfs or white dwarfs (WD) is very low. There have been reports of the discovery of exoplanets around horizontal \citep{Setia2010} and extreme-horizontal branch stars, that is, around helium (sdO) stars \citep{Bear2014} and  type B subdwarf (sdB) stars (\citealp{Silv2007,Geier2009,Charpinet2011,Silvotti2014}; 
\citealp[see the summary by][]{Heber2016}), but in most cases these planetary candidates were refuted in later papers \citep{Jones2014,Krzesinski2020,Krzesinski2015,Blo2019}. 
What remains are the two announcements by \cite{Silv2007} and \cite{Geier2009} which, as of yet, have no counterpart papers disputing the existence of the exoplanets.

In principle, planets orbiting an MS star at distances larger than the radius of the future red giant atmosphere can survive the red giant phase (RG) of their host \citep{Rasio1996}, but closer planets might not. According to the hydrodynamic simulations performed by \cite{Staff2016} for a 3.5 \(M_\odot\) zero-age main sequence (ZAMS) host, it takes less than $\sim$3 years for a 10 $M_{J}$ planet to spiral down onto the stellar core if the planet is engulfed by the stellar envelope during the RG stage. The time is longer (close to $\sim$100 years) if this happens during the asymptotic giant branch phase (AGB). 
The same fate awaits any exoplanets orbiting around close binary stars during their common envelope evolution. 

In this regard, the post-main-sequence evolution of the host star can be fatal for planets in the inner regions of the system. In contrast, the orbits of outer planets expand due to stellar mass-loss during the host RG and AGB phases \citep[see e.g.,][]{Veras2016a}. Therefore, we expect exoplanets at larger distances from the host than the former RG and AGB radii to survive, unless they migrate toward its host star later due to gravitational interactions with other planets in the system.  

In fact, observations of disintegrating planetesimals around the white dwarf WD~1145+017 \citep{Vander2018} suggest that there might be exoplanets in the system that are further away from the white dwarf, which perturb asteroid orbits causing the rocky material to fall onto the host star. However, even though WDs, due to their relatively low masses and small radii, are considered to be the best objects for searches of exoplanets around stars beyond the MS, until recently, no exoplanet has been found close to any WDs. The first indirect evidence for the presence of a planet around this type of star came from \cite{Gansicke2019}. 
The authors claim that they found a Neptune-mass exoplanet orbiting the white dwarf 
WD~J091405.30+191412.25 at $\sim$15 solar radii. The exoplanet was discovered by the 
analysis of stellar spectra from the Sloan Digital Sky Survey and follow-up spectroscopic observations. \cite{Gansicke2019} speculate that the exoplanet likely migrated toward its host star.
 
These two examples motivated us to investigate other evolved stars, including extreme 
horizontal branch stars (sdBs), WDs, as well as sdBs and WDs in binary systems. While 
WDs allow us to track the end-stages of exoplanetary evolution, sdBs provide us with 
the opportunity to track exoplanetary systems immediately after the host star RG phase. This is because these stars have lost most of the hydrogen from their envelopes before the helium flash and, therefore, they go directly to the WD cooling track after the RG phase, omitting the entire AGB stage \citep{Heber2016}. 

We are aware that our search might end up without any planet detections. However, a null result also has implications on the survival rate of planets through the host stars' late evolutionary stages and is scientifically interesting for our understanding of the long-term stability of planetary systems.

\section{Technique of analysis and time series data} 

\begin{table}
        \begin{center}
                \caption{List of objects}
                \label{tab1}
                {\scriptsize
                        \begin{tabular}{lcccc}\hline 
                                {\bf KIC ID\#} & {\bf Type } & {\bf Mass
                                        in \(M_\odot\) } & {\bf ref.} & {\bf remarks} \\ 
                                \hline
                                9472174 & sdBV+dM & $0.48 + 0.12$ & $1$ & eclips. binary,
                                $P$ = 0.12576528~d \\
                                7975824 & sdB+WD & $0.47 + 0.59$ & $2$ & eclips. binary,
                                $P$ = 0.40375026~d \\ 
                                8626021 & DBV &  0.55 & $3$ & single\\
                                \hline  
                        \end{tabular}
                }
        \end{center}
        \scriptsize{
                {\it ~~~~Notes:}
                ~~~$^{1}$\cite{Zola2013}, ~$^{2}$\cite{Bloem2011}, ~$^{3}$\cite{Giam2018} } 
        
\end{table}

In this work we compare three methods which can be used to analyze time series data. 
These are as follows: the Fourier transform (FT); eclipse timing variation diagrams 
based on the differences between observed times of minimum light (O) and those 
calculated from a linear ephemeris (C), referred to here as O$-$C (see \citealt{Ster2005}, and references therein or \citealt{Con2014} and \citealt{Bours2016}, for processes which can induce O$-$C variations); and finally time-delay methods \cite[][and references therein]{Shibahasi2012,Bal2014,Murphy2015}. 

While the first two are well known, the time-delay methods were developed to analyze delays in pulse times caused by the changing distance from the observer of a pulsating star in a binary or multiple system, using photometric data.
Here, we use the ``binarogram'' technique, introduced by \cite{Bal2014}. It allows for the efficient detection of the time-delay effect using the pulsation frequencies as a clock. 
This technique differs from other time-delay methods in that a plot (the binarogram) of the semi-major axis $a_{1} \sin(i)$ of the pulsating star's orbit around the center of mass ($i$ is the orbit inclination toward the observer) versus the orbital frequency is directly calculated from the stellar pulsations. The peaks (maxima) in the binarogram represent those orbital frequencies for which time-delay variations are present in the data. In other words, the binarogram is a periodogram of the orbital frequencies \citep{Bal2014}.

To calculate a binarogram from time series data of a pulsating star, one needs to do the following: determine the pulsation mode frequencies of the star; calculate the frequency phases within a data-window of a certain size (usually a few days); shift the window by a short stepping time (a couple of days); and recalculate the pulsation frequency phases. Having the phase variations as a function of time allows for the time-delay variation to be determined.
The method is very sensitive to variations in the pulsation amplitude and frequency, as well as to the interference of close neighboring pulsation modes which result in artifact orbital frequencies in the final binarogram. 

The method was successfully tested on $\delta$ Scuti stars, which are about 1.8 times as massive as the Sun \citep{Baglin1973}. The mass limit for an object which could be detected this way can be as low as $\sim$9\,$M_{J}$ with an orbital period of $\sim$800~d ($\sim$2~AU from the star). 
This is, of course, assuming that the observations allow us to achieve the limits of the method, which for long-cadence $\it{Kepler}$ data occurs when the time-delay amplitude is about 5\,s (or $a_{1}\sin(i)\,\approx$\,0.01\,AU, \citealp{Bal2014}). However, only a few of the claimed 
substellar mass candidates have been based on pulsation timings applied to $\delta$ Scuti stars \citep{Hermes2018}.

Since sdBs and WDs are less massive than $\delta$ Scuti stars, we should be able to detect exoplanetary objects of similar or lower masses.
\begin{figure}[!ht]     
        \centering
        \includegraphics[width=3.45in]{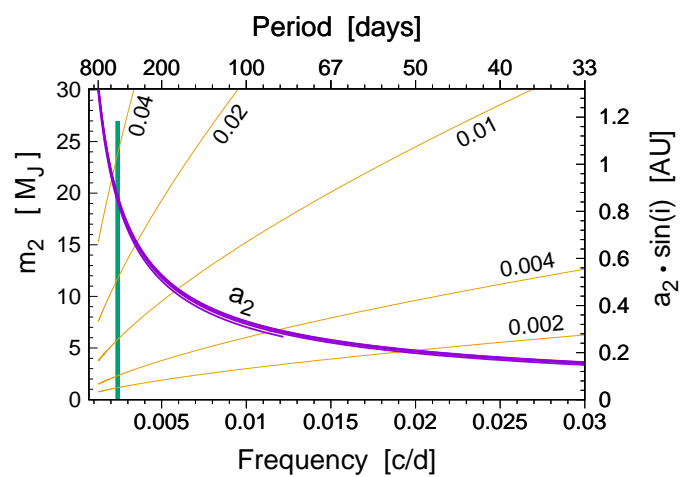}                 
        \caption{Exoplanet minimum mass $m_{2}$ as a function of orbital frequency and different
                values of $a_{1}\sin(i)$ (labeled in AU, orange lines) orbiting a typical sdB mass 
                $\sim$\,0.5\(M_\odot\). Violet lines represent the values (shown on the right) of 
                the semi-major axes $a_{2}\sin(i)$ of the exoplanetary orbit corresponding to 
                $a_{1}\sin(i)$. The top axis shows the exoplanet orbital period. The green line 
                marks the 0.0024\,c/d~frequency (referred to later in the text).
        }
        \label{fig_bg2}
        
\end{figure}
Assuming we have an exoplanet of mass $m_{2}$ in orbit around an $m_{1}$ mass star, we can plot $m_{2}$ as a function of its orbital frequency and $a_{1}\sin(i)$. In Fig.\,\ref{fig_bg2} we present the dependence of the minimum mass of $m_{2}$ (in Jupiter masses) on the orbital frequency for $a_{1}\sin(i)\,=$~0.002, 0.004, 0.01, and 0.02~AU and a 0.5\,\(M_\odot\) central $m_{1}$ stellar mass. The values of the semi-major axis $a_{2}\sin(i)$  of the exoplanet's orbit around the center of mass, which correspond to the above $a_{1}\sin(i)$ values, are shown in addition to the orbital frequency of the putative exoplanet, which is analyzed later in this work. 
As can be seen, for a typical mass of the sdB star and a 0.001--0.03~c/d frequency range, the time-delay method should allow for the detection of exoplanets of a minimum mass between 1--7\,$M_{J}$, for $a_{1}\sin(i)$ =\,0.02\,AU. 

Since all three detection methods rely on precise natural clocks, such as stellar pulsations or eclipses, our research was restricted to pulsating WDs, pulsating sdB (sdBV) stars, and short-period eclipsing binaries with WD or sdB components. We do not require the binary to have a pulsating star component; however, eclipsing binaries with pulsating components are the most suitable objects for this kind of study because they have two natural clocks, which can both be used to calculate binarograms. One of which is based on the binary orbital frequency and 
another one is based on the pulsation frequencies. 

In this part of our project, we explore photometric data from the primary part of the $\it{Kepler}$ telescope mission \citep{Boruc2010}, during which objects from the Kepler Input Catalog (KIC) \citep{Brown2011} were observed for up to four years. As such, their light curves allow for detailed studies of weak signals and longer orbital periods of exoplanets. All light curves were extracted from the short-cadence (SC) and long-cadence (LC) CCD Target Pixel Files, collected from the Barbara A. Mikulski Archive for Space Telescopes (MAST). The $\it{Kepler}$ data are divided into 90-day quarters (Q), and we use the same notation as \cite{Murphy2012} (for example ``LC Q\,3.2'' refers to long-cadence, quarter 3, month 2). Light curve extraction 
was done using the aperture reduction method, including the cotrending basis vectors (CBV) application (see the \textsc{kepprf}, \textsc{kepcotrend}, and \textsc{kepextract} task documentation of the \textsc{PyKE} command-line tools)\footnote{  https://keplerscience.arc.nasa.gov/software.html}. 
The extraction was performed by changing aperture sizes and CBV vector sets to maximize the light curve FT signal-to-noise ratios for the highest pulsation or orbital frequency amplitudes. 
Our approach results in an FT signal-amplitude increase of up to 20\% (for some months of the data) compared to the standard $\it{Kepler}$-pipeline light curve FT signal.

Finally, the extracted light curves were detrended for long-term flux variations using a moving average box with a width of 4--5 days, depending on the quarter of data. The minimum length of the detrending window comes from our simulations of short-period eclipsing binary light curves and O$-$C calculations using the methods described below. These simulations show that, for binary periods shorter than 1 day, a 3~day detrending window has a negligible impact on the O$-$C diagram. For an additional margin, we adopted 4~days as the minimum length of the detrending box in this work. This report outlines the results of the light curve analysis obtained for three objects listed in Table \ref{tab1}, that is, two eclipsing binaries and one helium white dwarf.

\section{KIC\,9472174: Light curves and O$-$C diagrams} 

To test and compare the methods, we chose the eclipsing binary KIC\,9472174, consisting of a red dwarf and sdBV stars (Table\,\ref{tab1}; \citealp[][]{Zola2013}) as the first target. The star has been intensively investigated by \cite{Ost_sdBdm2010} (radial velocity measurements, modeling, and variability of the sdBV component), \cite{Barl2012} (O$-$C diagrams, primary and secondary minima timings, and binary orbit eccentricity), \cite{Zola2013} (light curve modeling), 
and \cite{Bar2015} (O$-$C diagrams from prewhitened SC light curve, Jupiter-mass object detection, and asteroseismology). 
This $m_{Kep}$\,=\,12.26 mag ($\it{Kepler}$ filter magnitude) binary has two natural clocks, in the form of a pulsating sdBV star and mid-eclipse times, and both can be used to track time-delays. The claimed Jupiter-mass circumbinary companion orbiting the star at a distance of 0.92 AU with a period of 416 days \citep{Bar2015} makes the system even more valuable, since it allows us to test the sensitivity of the O$-$C and time-delay methods on the same object.

\subsection{Eclipse-time variations analysis} 
For our study, we used the Q\,5\,--\,Q\,17.2 SC data, which have already been analyzed by \cite{Bar2015}, as well as Q\,1\,--\,Q\,4 LC data omitted in previous analyses. Together, these sets of data are hereafter referred to as SC+LC.  
In order to cross-check the outcomes, we also took Q\,0\,--\,Q\,17.2 LC light curves from the Villanova catalog \cite[normalized flux\footnote{http://keplerebs.villanova.edu}]{Kirk2016}, 
hereafter LC~Vill, and processed it the same way as our SC+LC light curve that was extracted from pixel data (see below).

Because SC and LC $\it{Kepler}$ data are built up to produce 58.8~s and 29.4~min integrations from 6.02\,s exposures (Gilliland et al. 2010), respectively, the 0.12576-day orbital period \citep{Bar2015} of the binary is covered by $\sim$190 observing points in SC and $\sim$6 points in LC data. This translates into only $\sim$11--13 point coverage of each of the primary minima in the SC light curve, and practically zero or one point coverage of each of the minima in the LC light curve. This makes mid-eclipse time calculations using methods such as Kwee-van Woerden 
(KW) \citep{Kwee1956} vulnerable to any distortions of the shape of the minima, or even impossible in the case of the LC data. 

On the other hand, \cite{Bar2015} show that for SC data, the KW method gives results that are consistent with the other methods they used.
However, they obtained a smaller scatter in the O$-$C diagram when they determined the mid-eclipse times by modeling the light curve rather than minima fitting.
In our case, instead of a model light curve, we decided to use the template light curve fitting method to find times of minima \citep{Prib2012}. Both methods give a similar accuracy for a single minimum time determination. In this work, the template light curve is generated by phase-folding (over the orbital period) and averaging the phased light curve, using a moving average and a 100-point window. This single-period template is then fitted successively to the data to determine the mid-eclipse times.  

\begin{figure}
        \centering
        \includegraphics[width=3.4in, clip=]{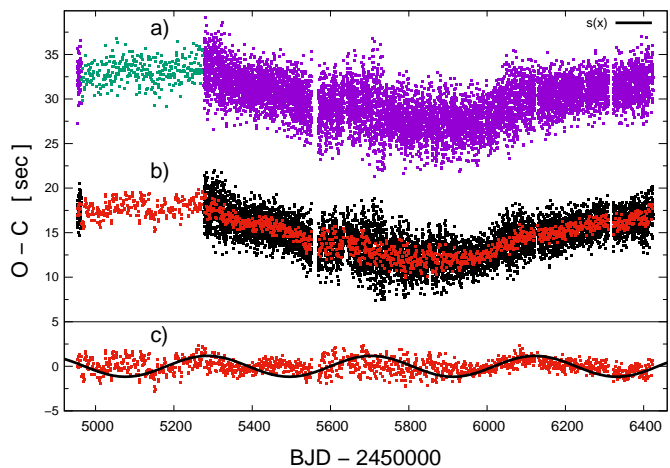}            
        \caption{KIC\,9472174: ~O$-$C diagrams calculated from: \textit{a)} SC and LC data (violet and green dots, respectively), and \textit{b)} data prewhitened from pulsations (black dots are the  O$-$C from the single-template fit to the SC data; red dots represent the O$-$C from the seven-period template fit to the SC+LC data). 
        The diagrams were shifted vertically to avoid overlapping. \newline
        Bottom panel \textit{c)}: O$-$C diagram from a seven-period template fit prewhitened for the 1430 day period. The black line shows a 416 day sinusoidal fit to the O$-$C from Fig.\,3 of \cite{Bar2015}. 
        }
        \label{fig1}
\end{figure}

\begin{figure}
        \centering 
        \includegraphics[width=3.4in, clip=]{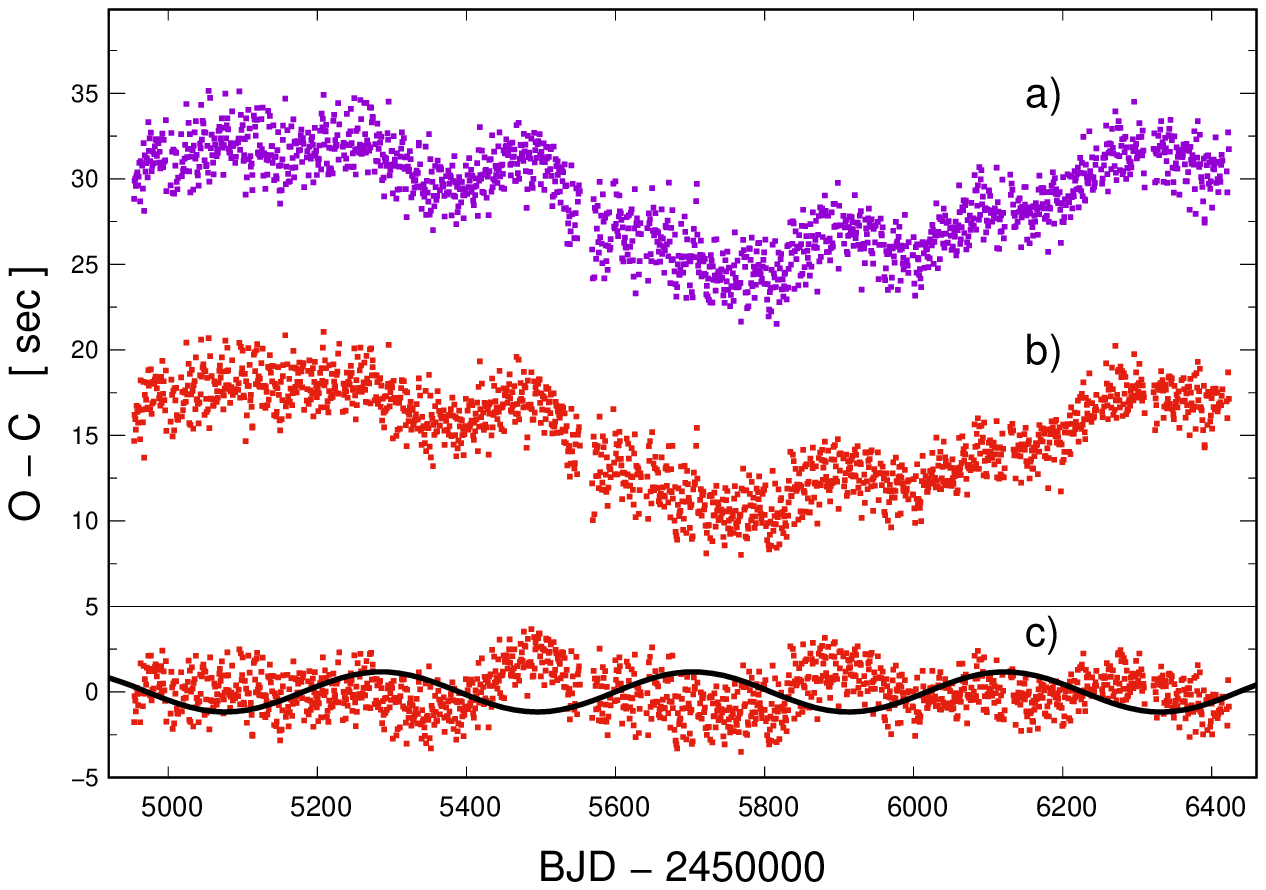}            
        \caption{KIC\,9472174: ~O$-$C diagrams calculated from seven-period template fits to \textit{a)}: the original LC~Vill data, and \textit{b)}: LC~Vill data 
        prewhitened from pulsations. The diagrams were shifted vertically for better visibility. \newline
        Bottom panel \textit{c)}: O$-$C shown at \textit{b)} prewhitened for the longest 1310 day period. The black line shows a 416 day sinusoidal fit to the O$-$C from Fig.\,3 of \cite{Bar2015}.  
        }
        \label{fig2}
\end{figure}

It turned out that the number of observations covering a single period of the KIC\,9472174 binary in the SC data was enough to fit a one-period long template to the light curve with an average accuracy of $\pm$4\,s. 
However, due to the poor coverage of the light curve by the observations in the LC data, these required a longer template covering several periods to achieve an adequate fit. After a number of trials, we found that the template generated from the LC data should be composed of at least seven repeated single-period templates. As such, the formal average fitting accuracy of the LC template was close to 50 seconds. As a result, from the template fitting, we obtained 8446~SC and 322~LC, and 1514 LC~Vill, eclipse-times and calculated O$-$C diagrams (Fig.\,\ref{fig1}\textit{a} and Fig.\,\ref{fig2}\textit{a}, respectively).
A single-period template fit to the SC light curve gives a similar ($\pm$3\,s) point 
scatter in the O$-$C diagram (Fig.\,\ref{fig1}\textit{a}), as is seen in the top panel of Fig.\,2 of \cite{Barl2012} and Fig.\,4 of \cite{Bar2015}. The scatter in the LC part (Fig.\,\ref{fig1}\textit{a}) is almost twice as small as the one observed in the SC part. 

Since the KIC\,9472174 binary system contains a variable sdB star, the pulsations of the sdBV component could be the main cause of the observed point scatter in the O$-$C diagram. To reduce this effect, the pulsations can be removed by "prewhitening" the light curves. To do so, following an initial fitting of the templates, the templates were subtracted from the SC+LC and LC~Vill data.
The residual light curves, which mosly consist of the sdBV pulsations, were prewhitened by simultaneous fitting (in amplitude, frequency, and phase) and subtraction from the residual light curves of identified pulsation modes. All modes with amplitudes higher than 0.02~ppt in the residual light curve were prewhitened. 

In the next step, the difference between the residual (after template subtraction) and prewhitened-residual light curves, which represents the pulsation signal in the original data, was subtracted from the original SC+LC and LC~Vill data. In this way, the SC+LC and LC~Vill light curves were cleaned of pulsations. The pulsation-free light curves were then again fitted with the templates and the new minima-times, as well as O$-$Cs, were calculated (Fig.\,\ref{fig1}\textit{b} and  Fig.\,\ref{fig2}\textit{b}). The overall SC O$-$C point scatter in 
Fig.\,\ref{fig1}\textit{b} became 30\% smaller than the one in Fig.\,4 from \cite{Bar2015}.

A further decrease in the point scatter in the SC part of the O$-$C diagram was achieved from a seven-period template fit to the SC pulsation-free data. The final O$-$C plot based on seven-period fits to both SC and LC light curves is presented 
in Fig.\,\ref{fig1}\textit{b}. 
All (Fig.\,\ref{fig1} and Fig.\,\ref{fig2}) diagrams were calculated for the  
BJD = 2454953.64324 + Epoch$\cdot$0.125765282 ephemeris. 

The bottom diagrams shown in Fig.\,\ref{fig1}\textit{c} and Fig.\,\ref{fig2}\textit{c} present the O$-$C diagrams from each panel \textit{b)}, detrended of the $\sim$1430 and $\sim$1310 day variations for the SC+LC and LC Vill data, respectively. In Fig.\,\ref{fig1}\textit{c} and Fig.\,\ref{fig2}\textit{c,}  the 416 day O$-$C variations given by \cite{Bar2015} in their Fig.\,3 O$-$C are shown; they attributed them to the presence of an exoplanet around KIC\,9472174.   
As we can see, the overall time course of the SC+LC O$-$C in Fig.\,\ref{fig1}\textit{c} mostly follows the black sinusoidal line, while the LC Vill O$-$C Fig.\,\ref{fig2}\textit{c} diagram differs drastically. 

\begin{figure}[!ht]
        \centering
        \includegraphics[width=3.4in]{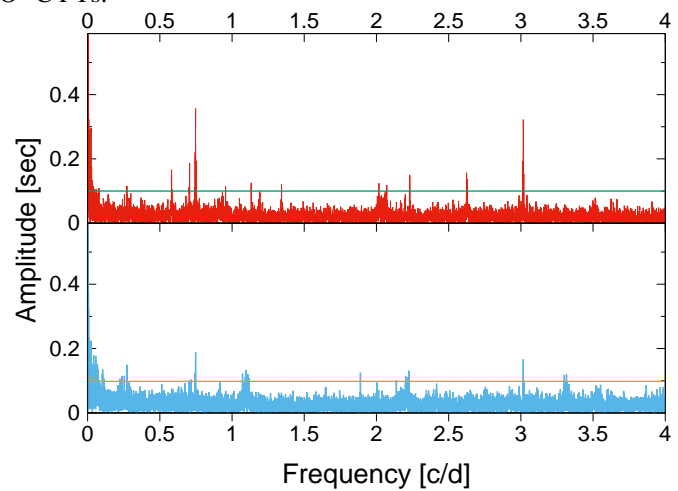}
        \caption{KIC\,9472174: ~~O$-$C FTs for SC \cite{Bar2015} O$-$C (top) and this work (bottom). 
        Horizontal lines mark 4$\sigma$ FT detection thresholds.}
        \label{fig3}
\end{figure}

To evaluate the differences, we compared FT amplitude spectra calculated for both our own O$-$C diagram and that from \cite{Bar2015}. Since the sampling rate for the SC part of the O$-$C is equal to the orbital period, the Nyquist frequency is $\sim$4~c/d (half of the orbital frequency  7.9513\,c/d). Consequently, the FT amplitude spectra were calculated between 0\,--\,4\,c/d and presented in Fig.\,\ref{fig3}. The \,$4\sigma$ FT detection thresholds shown in the figure were calculated between 0.2 and 4 c/d. They are nearly the same ($\sim$0.1\,seconds) for both our and the \cite{Bar2015} O$-$C FTs.

\begin{figure}[!ht]
        
        \centering
        \includegraphics[width=3.4in]{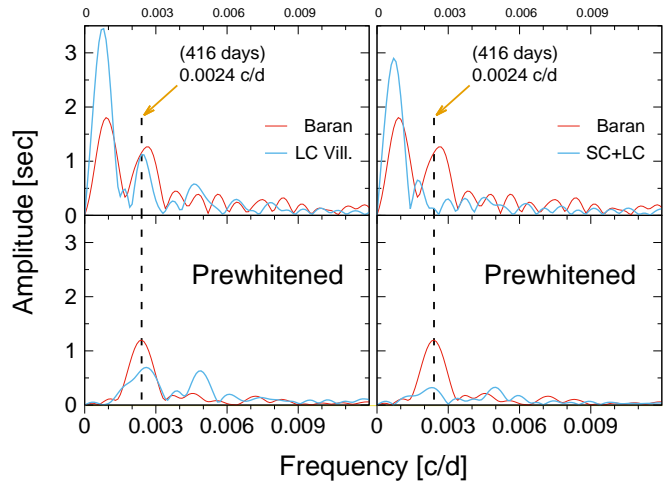}
        \caption{
                KIC\,9472174: ~~Enlargement of 0\,--\,0.012\,c/d FT-frequency region from Fig.\,\ref{fig1}. Red lines: SC \cite{Bar2015} data. Blue lines: This work, LC Villanova data (left panels), and SC+LC data (right panels). Top: The O$-$C FTs before prewhitening. Bottom: After prewhitening the O$-$Cs for the longest periods.
                Vertical dashed lines at 0.0024 c/d show the orbital frequency location of the putative exoplanet. 
        }
        \label{fig4}
\end{figure}   
   
The FT calculated for the O$-$C from \cite{Bar2015} has more high frequency peaks 
(for example at 0.6, 0.73, 0.75, 2.2, 2.6, and 3.0 c/d), which are reduced in amplitude or not present in our FT. Enlargements of the low (0\,--\,0.012\,c/d) frequency regions from Fig.\,\ref{fig3} are shown in Fig.\,\ref{fig4}. The FTs of the LC~Vill (left) and combined SC+LC (right) \text{O$-$Cs} in Fig.\,\ref{fig4} are over-plotted on top of the \cite{Bar2015} O$-$C diagram FTs. The bottom panels of Fig.\,\ref{fig4} present the FTs of the \text{O$-$Cs} that were prewhitened for the longest 1310 and 1430 day trends that are visible in the top panels at 0.000764\,c/d (LC Vill) and 0.000707\,c/d (SC+LC), respectively, 
as well as the 1114 day trend in \citep{Bar2015} O$-$C FT at 0.000898\,c/d. 
The orbital frequency of the putative planetary candidate from \cite{Bar2015} are 
marked at 0.0024\,c/d. The remaining highest-amplitude signals at 0.00256 $\pm$0.00002\,c/d and 0.00238 $\pm$0.00003\,c/d for LC~Vill and SC+LC O$-$Cs (bottom panels of Fig.\,\ref{fig4}), respectively, are close in frequencies to the 0.00240\,$\pm$0.00001\,c/d frequency found by \cite{Bar2015}. However, the 0.73\,$\pm$0.04\,s amplitude of the LC~Vill O$-$C signal is nearly twice as low as the 1.21\,$\pm$\,0.02\,s amplitude of the signal reported by \cite{Bar2015}, while the amplitude of SC+LC O$-$Cs 0.36\,$\pm$0.03\,s is beyond the 
FT detection threshold. 

This indicates a problem with previous interpretations of the signal visible at 
0.0024 c/d in the \cite{Bar2015} O$-$C. Our data analysis shows that the O$-$C 
amplitude variations are sensitive to the treatment of the $\it{Kepler}$ data 
before the O$-$C diagrams are calculated, and it suggests that the signal visible 
in the \cite{Bar2015} data is a consequence of the data extraction and processing.   
We notice that the $\sim$416 day periodicities are also observed in the amplitude
modulations of the pulsation modes (see \citealp{Bar2015} Fig.\,2 for the FT 
amplitude spectra of the KIC\,9472174 pulsation modes).
For some modes, their amplitude modulation can be extremely high and pulsations 
in these modes vanish periodically.
The amplitude modulation does not, by itself, have a direct impact on the 
O{\text-}C; however, when it occurs, the data reduction process or residual 
light curve processing may introduce changes to the shape of the light curve 
and thus affect the resulting measurements of the mid-eclipse times.

\subsection{Binarograms}

KIC\,9472174 has two natural clocks which can be used. One is the 7.9513\,c/d orbital frequency 
of the eclipsing binary itself, and the other is in the form of the pulsation modes 
of the sdBV component, which is determined from the FT amplitude spectrum of the 
residual light curve described above. 

\begin{figure}[!ht]
        \centering
        \includegraphics[width=3.5in]{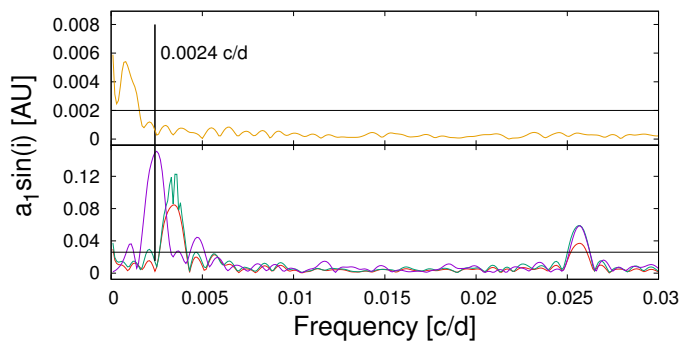}
        \caption{
                KIC\,9472174: ~Binarograms calculated using the orbital frequency of the 
                eclipsing binary (top panel, the vertical black line shows the orbital 
                frequency of the putative exoplanet) and using different pulsation 
                frequency sets of the sdBV component (bottom panel). Red represents 
                the set of 14 frequencies; green is the highest amplitude mode; and 
                violet is the second high amplitude mode.
        }
        \label{fig5}
\end{figure}

Using the orbital frequency of a binary to determine its time-delay variation is 
novel, but quite easy to do. We took the pulsation-free binary light curve (see 
above) and following the recipe given by \cite{Bal2014}, using a 4 day window 
and 2 day stepping time, we calculated the binarogram between 0\,--\,1\,c/d. 
Initially it had a lot of artifacts, which disappeared after we included the 
first harmonic (15.9026 c/d) of the orbital frequency in the calculation. 
The resulting binarogram is featureless; the only signal we found is located 
below 0.0015\,c/d and is due to the variable orbital frequency \citep[][Fig.\,8]{Bal2014}. 
For better visibility, we show a fraction of the orbital frequency binarogram between 
0\,--\,0.03\,c/d in Fig.\,\ref{fig5}.

The binarogram shows values of $a_{1}\sin(i)$, where $a_{1}$ is the length of the 
semi-major axis of the orbit of the binary component of mass $m_{1}$ around the 
center of mass of the system and a third body of mass $m_{3}$ ($i$ is the inclination 
toward the observer of the third body's orbit). Given that the orbital frequency of 
the third body derived from the binarogram is error-less, one solves the nonlinear 
least-squares problem \citep[][equations 1-4]{Bal2014} for $a_{1}\sin(i)$, and all 
of the uncertainty in the binarogram is attached to $a_{1}\sin(i)$. However, the 
estimated error for $a_{1}\sin(i)$ from the least-squares solution is the lowest 
uncertainty by far. The main difficulty with the binarogram arises when the frequency 
of the clock used to measure the time-delay is unstable, or if there is interference 
from different pulsation frequencies. Therefore, any noise in the binarogram is 
mainly due to these two effects. In practice, to find real signals in the binarogram, 
we use the same technique as for finding signals in the FT of the light curve, that 
is, by calculating the mean noise ($\sigma$) of the binarogram and setting the 
detection threshold at a certain $\sigma$ level. Here we set the detection threshold 
to 4\,$\sigma$. Since the mean noise in the orbital frequency binarogram is equal 
to 0.0005\,AU (calculated between 0\,--\,0.03\,c/d), the detection threshold is 
0.002\,AU (Fig.\,\ref{fig5}, top panel).

Now, if we consider $m_{1}$ and $m_{3}$ to be 0.6\,\(M_\odot\) (Table\,\ref{tab1}) 
and $\sim$0.002\,\(M_\odot\), respectively, and for the semi-major axis of the third 
body orbit to be $a_{3}$\,$\approx$\,0.9\,AU, the third body orbital frequency to 
be 0.0024\,c/d \citep{Bar2015}, and the orbital inclination of $m_{3}$ to be the 
same as the binary orbit of KIC\,9472174, that is, \text{\it i}\,$\approx$\,69$^\circ$ \citep{Bar2015}, the signal amplitude would be at $\sim$0.0028\,AU. Including the binarogram noise of 0.0005\,AU (see above), error propagation provides a lower limit of $\sim$0.0023\,AU for the signal amplitude. This is still above the 0.002\,AU binarogram detection threshold. Consequently, we cannot confirm the existence of an exoplanet on a 416 day orbit (at 0.9\,AU distance) around the binary system and with a minimum mass of 0.002\,\(M_\odot\), as claimed by 
\cite{Bar2015}.

\begin{figure}[!ht]
        
        \centering
        \includegraphics[width=3.5in]{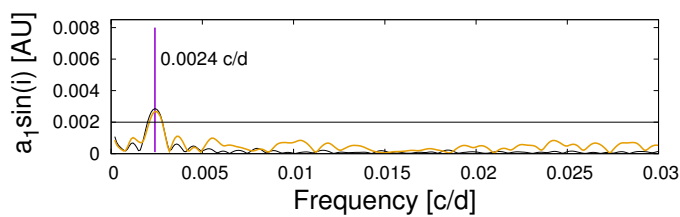}
        \caption{
                KIC\,9472174: ~Orbital frequency binarograms calculated from the synthetic light-cuves for the noiseless data (black line) and for the data with 3\,ppt noise added (orange line). Vertical (violet) and horizontal (black) lines show the orbital frequency of 
                the putative exoplanet and the 0.002\,AU detection threshold, respectively.
        }
        \label{fig5a}
\end{figure}

Since the actual $a_{1}\sin(i)$ error in the binarogram cannot be directly calculated, there might be some doubts whether a real exoplanetary signal, if it existed, would be detectable using this method on the data we have. Therefore, we prepared a synthetic light curve with induced time-delays from the putative exoplanet on a 416-day circular orbit around the binary at a distance of 0.9\,AU from the star and orbital inclination $i$\,=\,69$^\circ$ toward the observer. The flux from the binary was simulated by generating a sinusoidal light curve of the same frequency and amplitude as observed in the FT of the real binary light curve (i.e., 7.951319984\,c/d and 63 ppt, respectively), and a spacing between synthetic light curve points which exactly mimicked the time distribution of the observed points in the real data for KIC\,9472174. We then calculated orbital frequency binarograms as described above for both for the noiseless synthetic light curve Fig.\ref{fig5a} (black line) and with the addition of noise at the 3\,ppt level (same figure). This is more than the 1\,ppt standard deviation of the real data (which contains flux from the pulsating sdBV component) calculated from the ten-point moving average. 

As we can see for the noiseless synthetic light curve, the binarogram signal-amplitude $a_{1} \sin(i)$ and orbital frequency are reproduced exactly as introduced to the light curve (see the binarogram maximum at 0.0028\,AU and 0.0024 c/d in Fig.\ref{fig5a}). Adding noise to the synthetic light curve causes a signal-amplitude decrease from 0.0028 AU to 0.0027 AU, while the orbital frequency remains the same. The binarogram noise is also reproduced at the $\sim$0.0005\,AU level, as it is in the orbital frequency binarogram calculated from the real data (Fig.\,\ref{fig5}). As a result, the binarogram derived from synthetic data gives a detection threshold at the same level as observed in the real data. We also note that changes in the binary orbital frequency adopted for the binarogram calculations result in an artifact signal below 0.0015\,c/d, which is similar to the one observed in Fig.\,\ref{fig5}. This sets a $\sim$800-day limit on the longest orbital periods, which can be traced using the data we have. 
Taking the real and simulated data analysis into account, we can state that there is no exoplanet around the binary star KIC\,9472174  with the orbital parameters and minimum mass inferred by \citet{Bar2015}.  

Whether or not we can do the same thing using pulsation modes needs to be determined. 
As mentioned earlier, the main difficulty in calculating binarograms based on stellar 
pulsations is the interference between neighboring close pulsation frequencies and 
pulsation mode amplitude or frequency variations. The interference (or other pulsation mode variations) in frequencies result in binarogram artifacts, which must be distinguished from real signals. To do so, one needs to calculate binarograms for a few different sets of pulsation frequencies or data sets, for example by dividing the light curve into two or more parts and comparing the resulting binarograms with each other. The real signal is present at the same frequency and with a similar amplitude in all of the binarograms. This procedure requires a lot of computation time depending on the data length and the number of frequencies used. 
In the case of the SC $\it{Kepler}$ light curve, $\sim$1.5\,million points
and $\sim$20\,--\,30 pulsation modes, it takes about 24 hours to calculate a binarogram within a 1\,c/d orbital frequency range.  

In the bottom panel of Fig.\,\ref{fig5}, we present example binarograms calculated for the KIC\,9472174 SC data and three frequency sets: one consisting of 14 frequencies; another with the highest 195.7661\,c/d amplitude mode; and the last one with the second highest amplitude 40.0265\,c/d mode. As we can see, all signals below 0.005\,c/d differ in amplitude and frequencies and should be regarded as artifacts. Another signal at $\sim$0.026\,c/d has the same frequency for all three binarograms; however, for one set of frequencies, its amplitude is $\sim$40\% smaller. 
This suggests it is also not a real signal. 
As we mentioned earlier, the main difficulties in the pulsation-frequency binarograms 
are pulsation frequency variations and frequency interference, which cause an increase in binarogram noise and result in artifact signals. In this case, the detection threshold of the pulsation-frequency binarogram (bottom panel of Fig.\,\ref{fig5}) is $\sim$0.026\,AU. It is an order of magnitude greater than the one calculated for the orbital frequency binarogram, and we cannot reach the detectability required for the reliable detection of the putative exoplanet.          

In summary, the example of KIC\,9472174 shows that the data treatment can greatly 
influence the final shape of the O$-$C diagram and the conclusions drawn. However, 
the orbital frequency binarogram was sensitive enough to rule out the existence of 
the planetary companion suggested by \cite{Bar2015}. In fact, by using this 
binarogram, we could have detected a minimum $\sim$1.5\,$M_{J}$ mass object 
at a distance of 0.9\,AU from the binary on a 416-day orbit inclined at 
$i=69^\circ$, if it existed, or an $\sim$2\,$M_{J}$ mass object on orbit 
inclined at $i=43^\circ$.

\section{KIC\,7975824: O$-$C and binarogram}

The O$-$C analysis and binarogram of another eclipsing binary system composed of 
sdB and WD stars is presented in this section. The orbital period of this star is 
longer, 0.40375026 d (see Table\,\ref{tab1}), and the star is fainter 
($m_{Kep}$\,=\,14.66\,mag) than the previous one. The analysis was performed on 
joined Q\,1, Q\,5\,--\,Q12 SC, and Q13\,--\,Q\,17.2 LC data (SC+LC). The SC+LC 
light curves were obtained the same way as described above.  The fluxes and FT 
of the residual SC+LC light curve were calculated from single- and twenty-period 
template fits to the SC and LC light curves, respectively. The FT does not show 
any real periodic signals. The only features above the 0.008\,ppt FT detection 
threshold are some remnant signals around the orbital frequency and six known 
$\it{Kepler}$ artifacts at 440.43, 391.49, 342.55, 489.36, 685.11, and 
734.04 c/d (in order of decreasing amplitude). 
Therefore, we can confirm the finding by \cite{Bloem2011} that neither of the 
binary components are intrinsically variable down to this detection threshold.

\begin{figure}[!ht]
        
        \centering
        \includegraphics[width=3.4in]{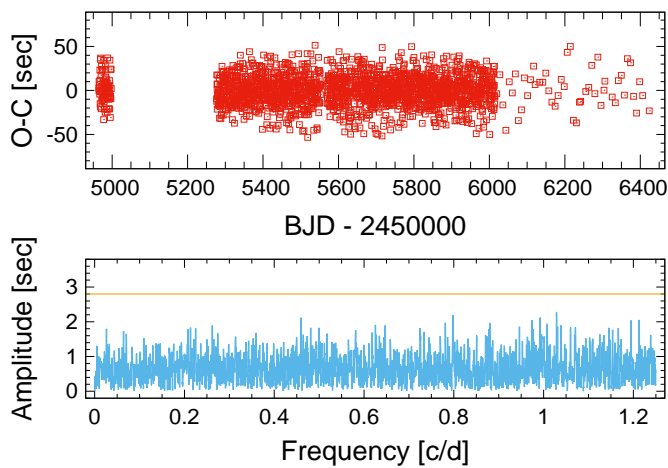}
        \caption{
                KIC\,7975824: ~O$-$C diagram (top) and its FT (bottom). The orange 
                line in the bottom panel marks the 4$\sigma$ FT detection threshold. 
        }
        \label{fig6}
\end{figure}

The mid-times of eclipses were calculated from the template fits to the light 
curve and the resulting O$-$C diagram is presented in the top panel of 
Fig.\,\ref{fig6}. The diagram was calculated for the ephemeris 
BJD\,=\,2454964.52672\,+\,Epoch\,$\cdot$\,0.403750221\,day, derived from 
a linear fit to the SC+LC minima times.

It can be seen that the points on the O$-$C diagram in the top panel of 
Fig.\,\ref{fig6} are scattered by about $\pm$\,50 seconds around zero 
(standard deviation $\pm$\,17 seconds). Open  squares at the end of the 
O$-$C diagram represent points obtained from a twenty-period template fit 
to the LC light curve. As far as we can tell, there are no variations in 
the diagram. In the bottom panel of the figure, we also show the O$-$C FT, 
which confirms that the orbital period was stable during the whole time of 
observation. 

\begin{figure}[!ht]
        
        \centering
        \includegraphics[width=3.5in]{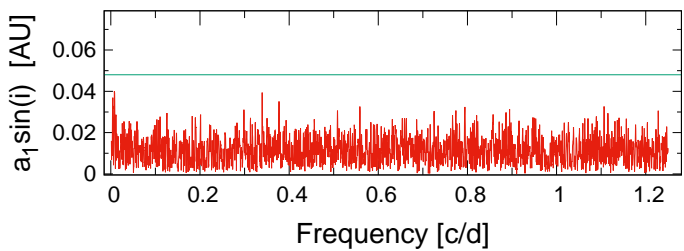}
        \caption{
                KIC\,7975824: ~Orbital frequency binarogram. The horizontal 
                line marks the 4$\sigma$ FT detection threshold.
        }
        \label{fig7}
\end{figure}

In the next step, we calculated the binarogram from the SC+LC light curve 
(Fig.\,\ref{fig7}), using the binary orbital frequency as a clock. Since 
the KIC\,797582 system is more massive (1.06\,\(M_\odot\), Table\,\ref{tab1}), 
we expect a lower sensitivity in the binarogram than was the case for the 
previous star. The binarogram detection threshold visible in Fig.\,\ref{fig7}
is about 0.048 AU, allowing for the detection of 0.05\,\(M_\odot\) bodies on 
a $\sim$1\,AU orbit around the system in the best conditions 
(\text{\it i}\,$\approx$\,90$^\circ$). 
The low sensitivity is apparently due to the higher mass of the system, 
the smaller amplitude of the light curve variations, and the larger noise 
in the data compared to the KIC\,9472174 light curve.

\section{KIC\,862602: Binarogram of a pulsating WD star}

KIC\,8626021 is a $m_{Kep}$\,=\,18.46\,mag pulsating helium atmosphere white
dwarf (DBV), showing about 12 pulsation frequencies in its light curve FT 
\citep{Zong2016,Giam2018}. This time, however, we have a single pulsating star 
and only the time-delay method can be used to search for substellar bodies 
orbiting the WD. As the mass of KIC\,8626021 is 0.55\,\(M_\odot\), for a 
0.002\,\(M_\odot\) object orbiting the DBV star at a distance of 1\,AU, 
one can therefore expect the binarogram signal to be $\sim$\,0.0036\,AU at best. 

\begin{figure}
        \centering
        \includegraphics[width=3.4in]{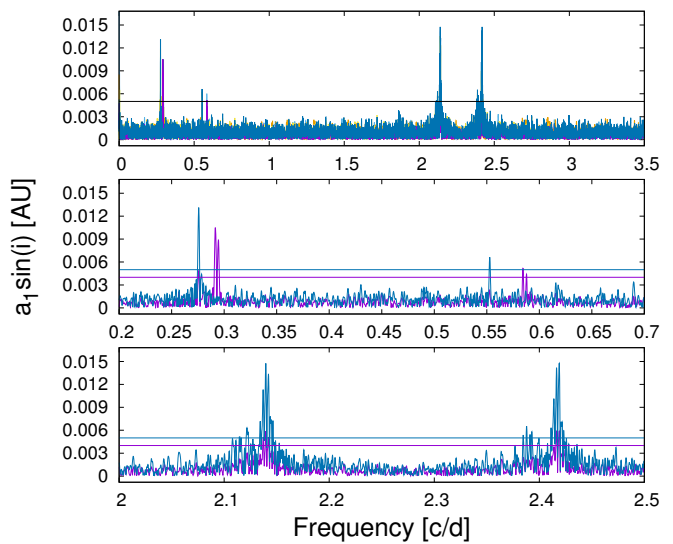}
        \caption{KIC\,8626021: Binarograms for different pulsation frequency sets 
        	(different line colors). Bottom panels: Enlargements of two artifact 
        	regions around 0.27\,c/d and 2.3\,c/d from the top panel. We note that 
        	the detection threshold changes (horizontal lines near $\sim$0.004\,AU) 
        	depending on the frequency sets used to calculate the binarograms.}
        \label{fig11}
\end{figure}

For the analysis, Q\,10\,--\,Q\,17.2 SC data were downloaded from MAST and 
the SC light curve was extracted as described in Section 2. The archive 
contains Q\,7 of LC, but it was not included since the LC data cannot 
follow the high frequency signals found in pulsating WDs.
For the binarogram calculation, the pulsation mode frequencies were taken 
from \cite{Zong2016}, who analyze the pulsation properties of the star in detail. 
However it turns out that, due to the variation of the amplitudes and phases 
of the pulsation frequencies and the presence of multiplets, the resulting 
binarograms are contaminated by a number of artifacts near 0.275, 0.29, 0.55, 0.58, 
2.15, and 2.4\,c/d (Fig.\,\ref{fig11}). We note that these binarograms were 
calculated for a wider orbital frequency range 0\,--\,3.5\,c/d  (Fig.\,\ref{fig11}, 
top panel) than for the previous two stars in order to demonstrate the issue, 
and the main part of our analysis consisted of ruling these artifacts out.

To do so, binarograms were calculated for different sets of frequencies as well 
as for the two halves of the data. Since a real $a_{1}\sin(i)$ signal should be 
independent of the subsets of data and frequency used for the calculations, the 
real signal should appear at the same frequency and amplitude in each binarogram,
with an accuracy corresponding to the binarogram resolution and its noise level. 
In our case, the binarogram noise is at $\sim$0.00125\,AU in Fig.\,\ref{fig11}; 
therefore, real signal amplitude variations can differ by approximately this 
amount for different subsets of the data. Different line colors in the figure 
represent binarograms calculated for different sets of pulsation frequencies. 
The regions around two pairs of strong signals near 0.4 and 2.25\,c/d of the 
top panel were enlarged in the frequency range and are presented in lower 
panels of Fig.\,\ref{fig11}. One can clearly see in the bottom panels that 
all the visible signal frequencies or amplitudes differ considerably; 
therefore, all of them can be classified as artifacts. 
The detection threshold drops down to 0.004\,AU for some sets of frequencies 
and this sets the minimum mass limit for a substellar object, which could be 
found at a distance of $\sim$1\,AU to about 2--3\,$M_{J}$, assuming $i=90^\circ$ 
(see Fig.\,\ref{fig_bg2}).

In summary, the low detection threshold of the pulsation-based binarogram allows us to search for  substellar objects of the lowest (0.002\,\(M_\odot\)) masses at a distance of 1 AU from a WD. The noise level can be surprisingly small, even for pulsation-mode binarograms and faint stars.  

\section{Conclusions}

Using time-delay, FT and O$-$C methods, we analyzed the light curves of two binaries and one pulsating WD. By comparing our O$-$C diagrams to those of \cite{Bar2015}, we find no evidence for the claimed Jupiter-mass planet at $\sim$0.9\, AU 
from the binary. We found that the $\sim$391 day cycle of variations observed in the LC Vill O$-$C (Fig.\,\ref{fig2}\textit{c}) is in antiphase to the 416 day variations visible in the \cite{Bar2015} O$-$C diagram, while variations in the SC+LC O$-$C diagram are below the detection threshold (Fig.\,\ref{fig4}). The shape of the final O$-$C depends on the data reduction techniques and light curve processing.   

The O$-$C diagram is not the definitive method of searching for orbiting companions to eclipsing binaries. Our final conclusion was drawn from the KIC\,9472174 binarogram (Fig.\,\ref{fig5}, top panel). This binarogram was based on the orbital frequency of the system and was sensitive enough to rule out the presence of a Jupiter- or higher-mass object on a 0.9\,AU and 416-day orbit (\textit{i}$\approx$69$^\circ$) around the star, such as is claimed by \cite{Bar2015}. The only possibility for our technique missing this planet would be an orbital inclination lower than $43^\circ$.
Our conclusion supplements the work by \cite{Pull2018} based on recent observations of seven sdB systems. The authors show that period variations of these stars cannot be explained simply on the basis of third bodies orbiting these systems. Out of seven binaries studied by authors, only three candidates support the circumbinary object hypothesis. 

We have also shown that the binarogram detection threshold depends on various factors, that is, light curve noise, an amplitude of light variations, and the stability of pulsation frequencies. However, the final detection threshold is difficult to predict. In the case of the bright eclipsing binary KIC\,9472174, the lowest noise level was achieved for the orbital-frequency binarogram; whereas, for the pulsation-mode binarogram, the noise was ten times higher. On the other hand, the noise level of the pulsation-frequency binarogram of the single WD, KIC\,8626021, which is a rather faint star, was comparable to that seen in the orbital-frequency binarogram of the binary KIC\,9472174. Additional problems are generated by artifacts. In the orbital frequency binarogram, the artifacts can be removed by the inclusion of orbital frequency harmonics (usually the first is enough) into the calculation. However, artifacts in pulsation-mode binarograms can only be distinguished from real signals by comparing many binarograms calculated for different subsets of frequencies or data. 

The methods we used are, in some cases, sensitive enough to search for giant planets around WDs, sdBs, and short period eclipsing binaries. While data from the primary part of the $\it{Kepler}$ telescope mission allows for these types of searches to be performed for longer $\sim$800-day orbital periods (i.e., larger, up to $\sim$1.4\,AU, distances from the stars) and $\sim$1\,$M_{J}$ exoplanets, the K2 SC light curves ($\sim$80 day photometric time-series data) restrict the search to short, 10\,--\,40 day periods and therefore to 7\,--\,20\,$M_{J}$ and higher mass objects.

\begin{acknowledgements} This work was supported by the National Science Centre,
        Poland. The project registration number is 2017/25/B/ST9/00879. The authors would like to thank Luis Balona from South African Astronomical Observatory, who kindly provided his software. 
\end{acknowledgements}


\bibliographystyle{aa}
\bibliography{38121corr}

\begin{thebibliography}{39}
\expandafter\ifx\csname natexlab\endcsname\relax\def\natexlab#1{#1}\fi

\bibitem[{{Baglin} {et~al.}(1973){Baglin}, {Breger}, {Chevalier}, {Hauck}, {Le
  Contel}, {Sareyan}, \& {Valtier}}]{Baglin1973}
{Baglin}, A., {Breger}, M., {Chevalier}, C., {et~al.} 1973, \aap, 23, 221

\bibitem[{{Balona}(2014)}]{Bal2014}
{Balona}, L.~A. 2014, \mnras, 443, 1946

\bibitem[{{Baran} {et~al.}(2015){Baran}, {Zola}, {Blokesz}, {{\O}stensen}, \&
  {Silvotti}}]{Bar2015}
{Baran}, A.~S., {Zola}, S., {Blokesz}, A., {{\O}stensen}, R.~H., \& {Silvotti},
  R. 2015, \aap, 577, A146

\bibitem[{{Barlow} {et~al.}(2012){Barlow}, {Wade}, \& {Liss}}]{Barl2012}
{Barlow}, B.~N., {Wade}, R.~A., \& {Liss}, S.~E. 2012, \apj, 753, 101

\bibitem[{{Bear} \& {Soker}(2014)}]{Bear2014}
{Bear}, E. \& {Soker}, N. 2014, \mnras, 437, 1400

\bibitem[{{Bloemen} {et~al.}(2011){Bloemen}, {Marsh}, {{\O}stensen},
  {Charpinet}, {Fontaine}, {Degroote}, {Heber}, {Kawaler}, {Aerts}, {Green},
  {Telting}, {Brassard}, {G{\"a}nsicke}, {Handler}, {Kurtz}, {Silvotti}, {Van
  Grootel}, {Lindberg}, {Pursimo}, {Wilson}, {Gilliland }, {Kjeldsen},
  {Christensen-Dalsgaard}, {Borucki}, {Koch}, {Jenkins}, \&
  {Klaus}}]{Bloem2011}
{Bloemen}, S., {Marsh}, T.~R., {{\O}stensen}, R.~H., {et~al.} 2011, \mnras,
  410, 1787

\bibitem[{{Blokesz} {et~al.}(2019){Blokesz}, {Krzesinski}, \&
  {Kedziora-Chudczer}}]{Blo2019}
{Blokesz}, A., {Krzesinski}, J., \& {Kedziora-Chudczer}, L. 2019, \aap, 627,
  A86

\bibitem[{{Borucki} {et~al.}(2010){Borucki}, {Koch}, {Basri}, {Batalha},
  {Brown}, {Caldwell}, {Caldwell}, {Christensen-Dalsgaard}, {Cochran},
  {DeVore}, {Dunham}, {Dupree}, {Gautier}, {Geary}, {Gilliland}, {Gould},
  {Howell}, {Jenkins}, {Kondo}, {Latham}, {Marcy}, {Meibom}, {Kjeldsen},
  {Lissauer}, {Monet}, {Morrison}, {Sasselov}, {Tarter}, {Boss}, {Brownlee},
  {Owen}, {Buzasi}, {Charbonneau}, {Doyle}, {Fortney}, {Ford}, {Holman},
  {Seager}, {Steffen}, {Welsh}, {Rowe}, {Anderson}, {Buchhave}, {Ciardi},
  {Walkowicz}, {Sherry}, {Horch}, {Isaacson}, {Everett}, {Fischer}, {Torres},
  {Johnson}, {Endl}, {MacQueen}, {Bryson}, {Dotson}, {Haas}, {Kolodziejczak},
  {Van Cleve}, {Chandrasekaran}, {Twicken}, {Quintana}, {Clarke}, {Allen},
  {Li}, {Wu}, {Tenenbaum}, {Verner}, {Bruhweiler}, {Barnes}, \&
  {Prsa}}]{Boruc2010}
{Borucki}, W.~J., {Koch}, D., {Basri}, G., {et~al.} 2010, Science, 327, 977

\bibitem[{{Bours} {et~al.}(2016){Bours}, {Marsh}, {Parsons}, {Dhillon},
  {Ashley}, {Bento}, {Breedt}, {Butterley}, {Caceres}, {Chote}, {Copperwheat},
  {Hardy}, {Hermes}, {Irawati}, {Kerry}, {Kilkenny}, {Littlefair},
  {McAllister}, {Rattanasoon}, {Sahman}, {Vu{\v{c}}kovi{\'c}}, \&
  {Wilson}}]{Bours2016}
{Bours}, M.~C.~P., {Marsh}, T.~R., {Parsons}, S.~G., {et~al.} 2016, \mnras,
  460, 3873

\bibitem[{{Brown} {et~al.}(2011){Brown}, {Latham}, {Everett}, \&
  {Esquerdo}}]{Brown2011}
{Brown}, T.~M., {Latham}, D.~W., {Everett}, M.~E., \& {Esquerdo}, G.~A. 2011,
  \aj, 142, 112

\bibitem[{{Cassan} {et~al.}(2012){Cassan}, {Kubas}, {Beaulieu}, {Dominik},
  {Horne}, {Greenhill}, {Wambsganss}, {Menzies}, {Williams}, {J{\o}rgensen},
  {Udalski}, {Bennett}, {Albrow}, {Batista}, {Brillant}, {Caldwell}, {Cole},
  {Coutures}, {Cook}, {Dieters}, {Dominis Prester}, {Donatowicz}, {Fouqu{\'e}},
  {Hill}, {Kains}, {Kane}, {Marquette}, {Martin}, {Pollard}, {Sahu}, {Vinter},
  {Warren}, {Watson}, {Zub}, {Sumi}, {Szyma{\'n}ski}, {Kubiak}, {Poleski},
  {Soszynski}, {Ulaczyk}, {Pietrzy{\'n}ski}, \& {Wyrzykowski}}]{Cassan2012}
{Cassan}, A., {Kubas}, D., {Beaulieu}, J.~P., {et~al.} 2012, \nat, 481, 167

\bibitem[{Charpinet {et~al.}(2011)Charpinet, Fontaine, Brassard, Green,
  Van~Grootel, Randall, Silvotti, Baran, {\O}stensen, Kawaler, \&
  Telting}]{Charpinet2011}
Charpinet, S., Fontaine, G., Brassard, P., {et~al.} 2011, \nat, 480, 496

\bibitem[{{Conroy} {et~al.}(2014){Conroy}, {Pr{\v{s}}a}, {Stassun}, {Orosz},
  {Fabrycky}, \& {Welsh}}]{Con2014}
{Conroy}, K.~E., {Pr{\v{s}}a}, A., {Stassun}, K.~G., {et~al.} 2014, \aj, 147,
  45

\bibitem[{{G{\"a}nsicke} {et~al.}(2019){G{\"a}nsicke}, {Schreiber}, {Toloza},
  {Fusillo}, {Koester}, \& {Manser}}]{Gansicke2019}
{G{\"a}nsicke}, B.~T., {Schreiber}, M.~R., {Toloza}, O., {et~al.} 2019, \nat,
  576, 61

\bibitem[{{Geier} {et~al.}(2009){Geier}, {Edelmann}, {Heber}, \&
  {Morales-Rueda}}]{Geier2009}
{Geier}, S., {Edelmann}, H., {Heber}, U., \& {Morales-Rueda}, L. 2009, \apjl,
  702, L96

\bibitem[{{Giammichele} {et~al.}(2018){Giammichele}, {Charpinet}, {Fontaine},
  {Brassard}, {Green}, {Van Grootel}, {Bergeron}, {Zong}, \&
  {Dupret}}]{Giam2018}
{Giammichele}, N., {Charpinet}, S., {Fontaine}, G., {et~al.} 2018, \nat, 554,
  73

\bibitem[{Heber(2016)}]{Heber2016}
Heber, U. 2016, \pasp, 128, 082001

\bibitem[{{Hermes}(2018)}]{Hermes2018}
{Hermes}, J.~J. 2018, {Timing by Stellar Pulsations as an Exoplanet Discovery
  Method}, ed. H.~Deeg \& J.~Belmonte, Handbook of Exoplanets. Springer, Cham.,
  787

\bibitem[{{Jones} \& {Jenkins}(2014)}]{Jones2014}
{Jones}, M.~I. \& {Jenkins}, J.~S. 2014, \aap, 562, A129

\bibitem[{{Kirk} {et~al.}(2016){Kirk}, {Conroy}, {Pr{\v{s}}a}, {Abdul-Masih},
  {Kochoska}, {Matijevi{\v{c}}}, {Hambleton}, {Barclay}, {Bloemen}, {Boyajian},
  {Doyle}, {Fulton}, {Hoekstra}, {Jek}, {Kane}, {Kostov}, {Latham}, {Mazeh},
  {Orosz}, {Pepper}, {Quarles}, {Ragozzine}, {Shporer}, {Southworth},
  {Stassun}, {Thompson}, {Welsh}, {Agol}, {Derekas}, {Devor}, {Fischer},
  {Green}, {Gropp}, {Jacobs}, {Johnston}, {LaCourse}, {Saetre}, {Schwengeler},
  {Toczyski}, {Werner}, {Garrett}, {Gore}, {Martinez}, {Spitzer}, {Stevick},
  {Thomadis}, {Vrijmoet}, {Yenawine}, {Batalha}, \& {Borucki}}]{Kirk2016}
{Kirk}, B., {Conroy}, K., {Pr{\v{s}}a}, A., {et~al.} 2016, \aj, 151, 68

\bibitem[{{Krzesinski}(2015)}]{Krzesinski2015}
{Krzesinski}, J. 2015, \aap, 581, A7

\bibitem[{{Krzesinski} {et~al.}(2020){Krzesinski}, {Blokesz}, {Og{\l}oza}, \&
  {Dr{\'o}{\.z}d{\.z}}}]{Krzesinski2020}
{Krzesinski}, J., {Blokesz}, A., {Og{\l}oza}, W., \& {Dr{\'o}{\.z}d{\.z}}, M.
  2020, in IAU Symposium, ed. B.~G. {Elmegreen}, L.~V. {T{\'o}th}, \&
  M.~{G{\"u}del}, Vol. 345, 306--307

\bibitem[{{Kwee} \& {van Woerden}(1956)}]{Kwee1956}
{Kwee}, K.~K. \& {van Woerden}, H. 1956, \bain, 12, 327

\bibitem[{{Murphy}(2012)}]{Murphy2012}
{Murphy}, S.~J. 2012, \mnras, 422, 665

\bibitem[{{Murphy} \& {Shibahashi}(2015)}]{Murphy2015}
{Murphy}, S.~J. \& {Shibahashi}, H. 2015, \mnras, 450, 4475

\bibitem[{{{\O}stensen} {et~al.}(2010){{\O}stensen}, {Green}, {Bloemen},
  {Marsh}, {Laird}, {Morris}, {Moriyama}, {Oreiro}, {Reed}, {Kawaler}, {Aerts},
  {Vu{\v{c}}kovi{\'c}}, {Degroote}, {Telting}, {Kjeldsen}, {Gilliland},
  {Christensen-Dalsgaard}, {Borucki}, \& {Koch}}]{Ost_sdBdm2010}
{{\O}stensen}, R.~H., {Green}, E.~M., {Bloemen}, S., {et~al.} 2010, \mnras,
  408, L51

\bibitem[{{Pribulla } {et~al.}(2012){Pribulla }, {Va{\v{n}}ko}, {Ammler-von
  Eiff}, {Andreev}, {Aslant{\"u}rk}, {Awadalla}, {Balu{\v{d}}ansk{\'y}},
  {Bonanno}, {Bo{\v{z}}i{\'c}}, {Catanzaro}, {{\c{C}}elik}, {Christopoulou},
  {Covino}, {Cusano}, {Dimitrov}, {Dubovsk{\'y}}, {Eigmueller}, {Esmer},
  {Frasca}, {Hamb{\'a}lek}, {Hanna}, {Hanslmeier}, {Kalomeni}, {Kjurkchieva},
  {Krushevska}, {Kudzej}, {Kundra}, {Kuznyetsova}, {Lee}, {Leitzinger},
  {Maciejewski}, {Moldovan}, {Morais}, {Mugrauer}, {Neuh{\"a}user},
  {Niedzielski}, {Odert}, {Ohlert}, {{\"O}zavc{\i}}, {Papageorgiou},
  {Parimucha}, {Poddan{\'y}}, {Pop}, {Raetz}, {Raetz}, {Romanyuk},
  {Ru{\v{z}}djak}, {Schulz}, {{\c{S}}enavc{\i}}, {Srdoc}, {Szalai},
  {Sz{\'e}kely}, {Sudar}, {Tezcan}, {T{\"o}r{\"u}n}, {Turcu}, {Vince}, \&
  {Zejda}}]{Prib2012}
{Pribulla }, T., {Va{\v{n}}ko}, M., {Ammler-von Eiff}, M., {et~al.} 2012,
  Astron. Nachr., 333, 754

\bibitem[{{Pulley} {et~al.}(2018){Pulley}, {Faillace}, {Smith}, {Watkins}, \&
  {von Harrach}}]{Pull2018}
{Pulley}, D., {Faillace}, G., {Smith}, D., {Watkins}, A., \& {von Harrach}, S.
  2018, \aap, 611, A48

\bibitem[{{Rasio} {et~al.}(1996){Rasio}, {Tout}, {Lubow}, \&
  {Livio}}]{Rasio1996}
{Rasio}, F.~A., {Tout}, C.~A., {Lubow}, S.~H., \& {Livio}, M. 1996, \apj, 470,
  1187

\bibitem[{{Setiawan} {et~al.}(2010){Setiawan}, {Klement}, {Henning}, {Rix},
  {Rochau}, {Rodmann}, \& {Schulze-Hartung}}]{Setia2010}
{Setiawan}, J., {Klement}, R.~J., {Henning}, T., {et~al.} 2010, Science, 330,
  1642

\bibitem[{{Shibahashi} \& {Kurtz}(2012)}]{Shibahasi2012}
{Shibahashi}, H. \& {Kurtz}, D.~W. 2012, \mnras, 422, 738

\bibitem[{Silvotti {et~al.}(2014)Silvotti, Charpinet, Green, Fontaine, Telting,
  {\O}stensen, Van~Grootel, Baran, Schuh, \& Fox~Machado}]{Silvotti2014}
Silvotti, R., Charpinet, S., Green, E., {et~al.} 2014, \aap, 570

\bibitem[{{Silvotti} {et~al.}(2007){Silvotti}, {Schuh}, {Janulis}, {Bernabei},
  {{\O}stensen}, {Solheim}, {Bruni}, {Gualand i}, {Oswalt}, {Bonanno}, \&
  {Mignemi}}]{Silv2007}
{Silvotti}, R., {Schuh}, S., {Janulis}, R., {et~al.} 2007, Astronomical Society
  of the Pacific Conference Series, Vol. 372, {The O--C Diagram of the Subdwarf
  B Pulsating Star HS 2201+2610: Detection of a Giant Planet?}, ed.
  R.~{Napiwotzki} \& M.~R. {Burleigh}, 369

\bibitem[{Staff {et~al.}(2016)Staff, De~Marco, Wood, Galaviz, \&
  Passy}]{Staff2016}
Staff, J., De~Marco, O., Wood, P., Galaviz, P., \& Passy, J.-C. 2016, \mnras,
  458, 832

\bibitem[{{Sterken}(2005)}]{Ster2005}
{Sterken}, C. 2005, Astronomical Society of the Pacific Conference Series, Vol.
  335, {The O-C Diagram: Basic Procedures}, ed. C.~{Sterken}, 3

\bibitem[{{Vanderburg} \& {Rappaport}(2018)}]{Vander2018}
{Vanderburg}, A. \& {Rappaport}, S.~A. 2018, {Transiting Disintegrating
  Planetary Debris Around WD 1145+017}, ed. H.~Deeg \& J.~Belmonte, Handbook of
  Exoplanets. Springer, Cham., 2603

\bibitem[{{Veras} {et~al.}(2016){Veras}, {Mustill}, {G{\"a}nsicke}, {Redfield},
  {Georgakarakos}, {Bowler}, \& {Lloyd}}]{Veras2016a}
{Veras}, D., {Mustill}, A.~J., {G{\"a}nsicke}, B.~T., {et~al.} 2016, \mnras,
  458, 3942

\bibitem[{{Zola} \& {Baran}(2013)}]{Zola2013}
{Zola}, S. \& {Baran}, A. 2013, Central European Astrophysical Bulletin, 37,
  227

\bibitem[{{Zong} {et~al.}(2016){Zong}, {Charpinet}, {Vauclair}, {Giammichele},
  \& {Van Grootel}}]{Zong2016}
{Zong}, W., {Charpinet}, S., {Vauclair}, G., {Giammichele}, N., \& {Van
  Grootel}, V. 2016, \aap, 585, A22

\end{thebibliography}

\end{document}